\documentclass[prl,twocolumn,showpacs,preprintnumbers]{revtex4-1}
\usepackage{graphicx}
\usepackage{amsmath}
\usepackage{amssymb}
\usepackage{times}
\usepackage{color}
\usepackage{ulem}
\usepackage{physics}
\usepackage{cleveref}
\usepackage{xr}
\usepackage{here}

\def\D{\Delta}

\def\G{\Gamma}

\def\s{\sigma}

\def\w{\omega}
\def\ua{\uparrow}
\def\da{\downarrow}

\def\Vec#1{\mathbf #1}

\newcommand{\Tc}{T_\text{c}}

\newcommand{\nb}{\bar{n}}

\begin{document}

\title{Exotic pairing state in quasicrystalline superconductors under magnetic field}

\author{Shiro Sakai$^1$ and Ryotaro Arita$^{1,2}$}

\affiliation{
$^1$Center for Emergent Matter Science, RIKEN, Wako, Saitama 351-0198, Japan\\
$^2$Department of Applied Physics, University of Tokyo, Hongo, Tokyo 113-8656, Japan
}
\date{\today}
\begin{abstract}
We theoretically study the effect of a magnetic field on quasicrystalline superconductors, by modelling them as the attractive Hubbard model on the Penrose-tiling structure. We find that at low temperatures and under a high magnetic field there appears an exotic superconducting state with the order parameter changing its sign in real space. We discuss the state in comparison with the Fulde-Ferrell-Larkin-Ovchinnikov state proposed many years ago for periodic systems, clarifying commonalities and differences. It is remarkable that, even in the absence of periodicity, the electronic system finds a way to keep a coherent superconducting state with a spatially sign-changing order parameter compatible with the underlying quasiperiodic structure. 
\end{abstract}
\maketitle

Quasicrystals \cite{Shechtman84} are a platform of novel electronic properties because of their underlying fractal crystalline structure without periodicity.
Unlike the conventional periodic crystals, quasicrystals have no well-defined momentum space (with respect to the relative coordinate) and hence no Fermi surface even though many of them show metallic behaviors.
Early theoretical studies on related quasiperiodic structures revealed various nontrivial electronic properties, such as a presence of a confined state \cite{kohmoto86PRL,arai88}, singular-continuous spectrum \cite{kohmoto83,ostlund83,tsunetsugu86} and multifractal dimensions \cite{sutherland86,tokihiro88}.
While these studies concerned noninteracting electrons, increasingly more attentions have been paid to electron-correlation problems in the quasiperiodic systems \cite{watanabe13,shaginyan13,andrade15,thiem15,takemori15,takemura15,hartman16,watanabe16,otsuki16,shinzaki16,koga17} since the discovery of a quantum critical behavior in Au$_{51}$Al$_{34}$Yb$_{15}$ quasicrystal \cite{deguchi12}.

Another interesting observation in recent experiments is the discovery of superconductivity in Al-Zn-Mg quasicrystalline alloy \cite{kamiya18}.
This was the first observation of an electronic long-range order in quasicrystals.
In parallel with this experimental search for quasicrystalline superconductors,  we have theoretically studied a possible superconductivity on the Penrose-tiling structure \cite{penrose74}, which is a prototype of quasicrystalline structures. 
In Ref.~\onlinecite{sakai17}, we found that, even without a Fermi surface, there exists a superconducting state with spatially extended Cooper pairs in a weak-coupling regime of the attractive Hubbard model.
This Cooper pairing cannot be straightforwardly captured by Bardeen-Cooper-Schrieffer picture in the sense that the latter is based on a description in a momentum space absent in quasiperiodic systems.
Instead, we may expect novel superconducting properties resulting from an interplay between the macroscopic coherence of the superconductivity and a fractal geometry of the quasiperiodic lattice.

\begin{figure}[tb]
\center{
\includegraphics[width=0.48\textwidth]{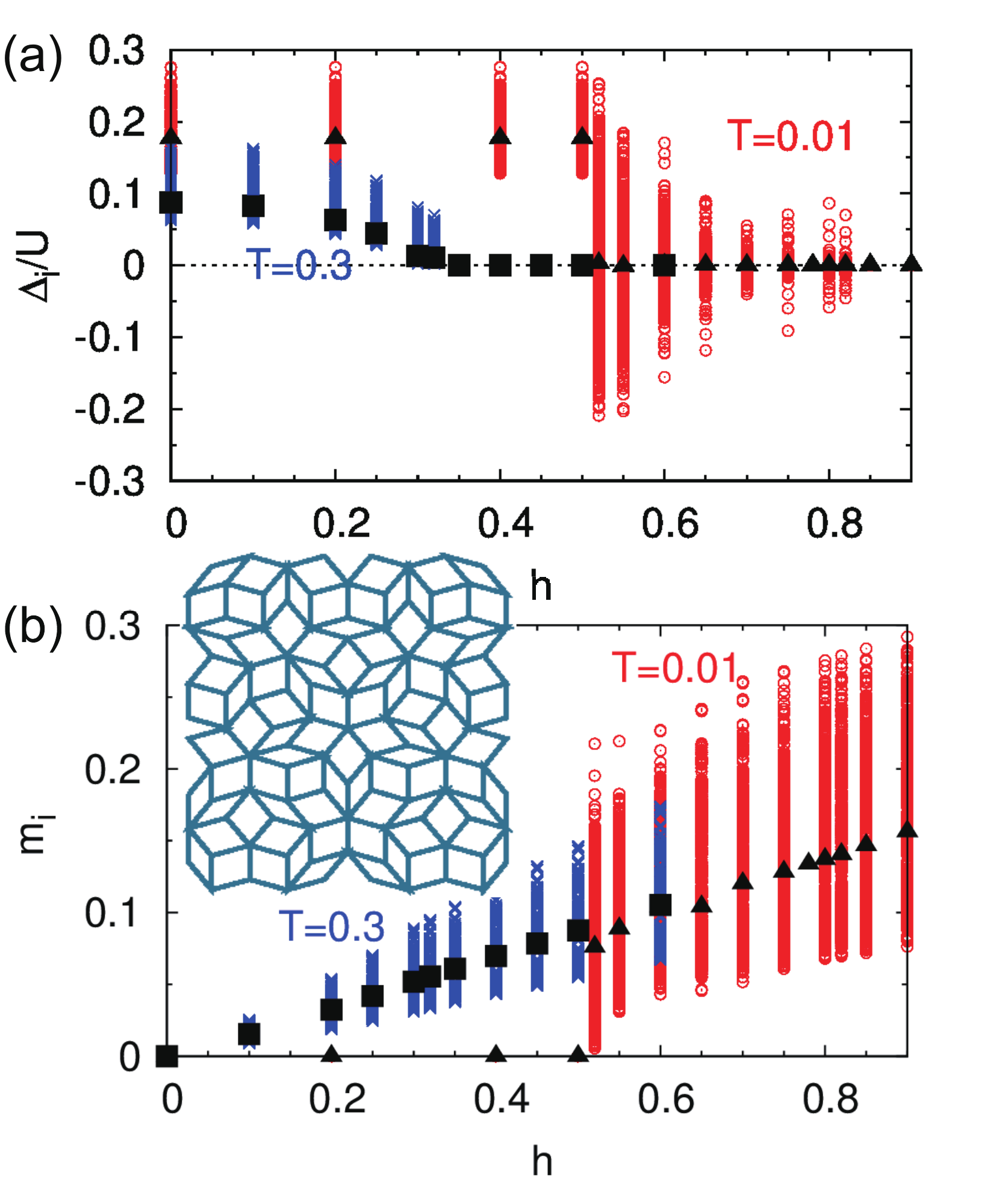}}
\caption{(a) Superconducting order parameter $\D_i/U$ plotted against the magnetic-field strength $h$ at $T=0.01$ (red circles) and $0.3$ (blue crosses). $U=-3$, $\nb=0.5$ and $N=11006$ are used. Black triangles and squares denote the values averaged over the sites at $T=0.01$ and $0.3$, respectively.
(b) Corresponding plot of the magnetization $m_i$. Inset shows an example of the Penrose tiling. 
}
\label{fig:op-h}
\end{figure}

\begin{figure*}[tb]
\center{
\includegraphics[width=\textwidth]{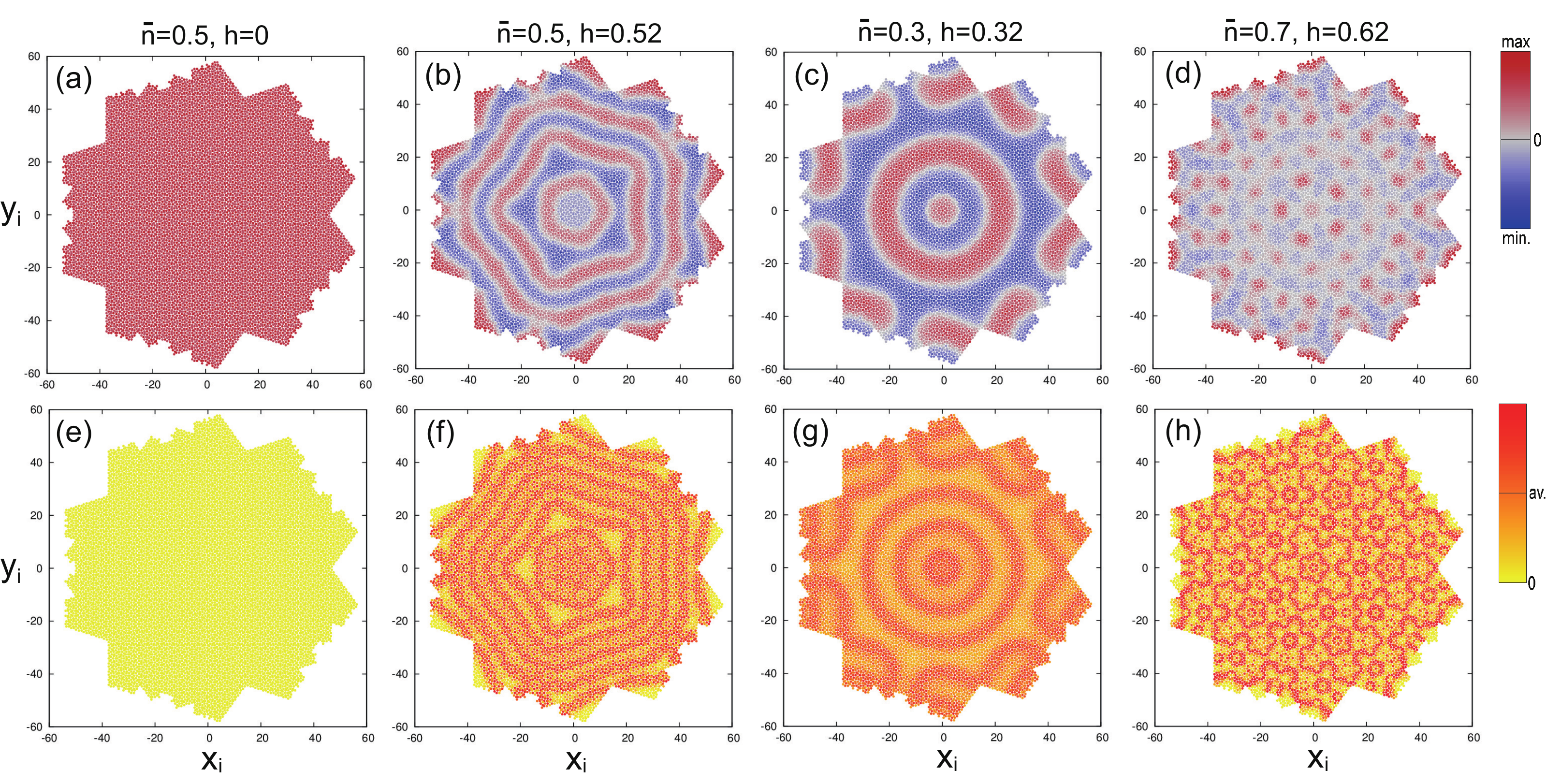}}
\caption{Spatial pattern of (a-d) the superconducting order parameter $\D_i$ and (e-h) magnetization $m_i$ for $T=0.01$, $U=-3$ and $N=11006$.
(a)(e) $\nb=0.5, h=0$, (b)(f) $\nb=0.5, h=0.52$, (c)(g) $\nb=0.3, h=0.32$, and (d)(h) $\nb=0.7, h=0.62$.}
\label{fig:op-site}
\end{figure*}

In this Letter, we study the effect of magnetic field on quasiperiodic superconductors. 
In a high-field region just below the critical field, we find a novel superconducting state with a spatially alternating sign of the order parameter.
The state reminds us of an inhomogeneous superconducting state, called Fulde-Ferrell-Larkin-Ovchinnikov (FFLO) state  \cite{fulde64,larkin65,casalbuoni04,matsuda07}, in periodic systems:
In particular, it may be analogous to the Larkin-Ovchinnikov state with the alternating-sign superconducting order parameter $\D(\Vec{r})=\D_0 \cos(\Vec{q}\cdot\Vec{r})$ \cite{larkin65} characterized by a spin-dependent momentum shift $\Vec{q}$ of the normal-state Fermi surfaces.
However, the alternating-sign structure which we found possesses a five-fold rotational symmetry complying with the geometry of the Penrose tiling, and is not characterized by any $\Vec{q}$ vector unlike the FFLO state.
The emergence of this exotic superconductivity is remarkable when we consider that the FFLO state is unstable against impurities \cite{aslamazov69,agterberg01}: Despite a naive expectation that the quasiperiodic structure would work as a random potential and thereby easily destroy such an alternating-sign pairing, the electronic system finds a way to keep a coherent alternating-sign state by organizing itself to comply with the underlying quasiperiodic structure.

Our model is based on the attractive Hubbard model on a Penrose tiling \cite{penrose74}, introduced in Ref.~\onlinecite{sakai17} as a simple theoretical model to discuss a superconductivity in quasiperiodic systems.
The Penrose tiling [Inset to Fig.~\ref{fig:op-h} (b)] is a prototypical quasicrystalline structure, covering an entire plane by only two different types of rhombuses. 
We regard each vertex of the rhombuses as a site, and put an electron hopping $t$ between two sites connected by the edge of the rhombuses.
We mainly consider an open-boundary cluster of $N=11006$ sites (results for $N=4181$ sites are presented in Supplementary Material) generated by iteratively applying the inflation-deflation rule \cite{levine84}.

Under an external magnetic field, the Hamiltonian reads 
\begin{align}
\hat{H}&=-t \sum_{\langle ij\rangle\s} \hat{c}_{i\s}^\dagger \hat{c}_{j\s}^{\phantom {\dagger}}- \mu\sum_{i\s}\hat{n}_{i\s}
+U\sum_{i}\hat{n}_{i\ua} \hat{n}_{i\da} \nonumber\\
&-h \sum_{i} (\hat{n}_{i\ua}-\hat{n}_{i\da}), 
\label{hubbard}
\end{align}
where $\hat{c}_{i\s}$ ($\hat{c}_{i\s}^\dagger$) annihilates (creates) an electron of spin $\s (=\ua, \da)$ at site $i$ and $\hat{n}_{i\s}\equiv \hat{c}_{i\s}^\dagger \hat{c}_{i\s}$.
The chemical potential $\mu$ is determined self-consistently to reproduce a given average electron density $\nb\equiv \frac{1}{N}\sum_{i\s} \langle \hat{n}_{i\s}\rangle$. In most cases, we set $\nb <1$ to avoid any possible peculiarity originating from the confined states \cite{kohmoto86PRL,arai88}.
We consider an onsite attraction $U<0$ to discuss the superconductivity.
The last term represents the Zeeman effect. 
Here we have neglected the orbital motion of the electrons by considering a magnetic field applied in parallel with the tiling plane.
We hereafter set $t=1$ as the unit of energy.

At zero magnetic field, the model (\ref{hubbard}) shows a superconductivity in a wide range of the parameters $U<0$ and $\nb$ at low temperatures, while the character of the superconductivity substantially changes with these parameters \cite{sakai17}.
In this study, we focus on a weak-coupling regime, where the Cooper pairs are spatially extended (despite the inapplicability of momentum-space picture of the pairing) and hence would be strongly influenced by the underlying quasiperiodic geometry.
In addition, this weak-coupling regime would be most relevant to a real quasicrystal \cite{kamiya18}. 
The spatially extended nature, however, requires a large-size numerical simulation to obtain well size-converged results.
For this reason and since the Cooper pairs extend less for a larger $|U|$, we choose a moderately weak value $U=-3$ in the following (we discuss $U$ dependence in Fig.~\ref{fig:pdu} of Supplementary Material).

Based on the model (\ref{hubbard}), we have numerically solved the Bogoliubov - de Gennes (BdG) equation,
\begin{align}
&E_{m\s}\left( \begin{array}{c}
                     u_{mi\s}\\
                     v_{mi\bar{\s}}
                   \end{array}\right)
=-t\sum_{j\in \{ {\rm n.n.\ of}\ i\}}\left( \begin{array}{c}
                     u_{mj\s}\\
                     -v_{mj\bar{\s}}
                   \end{array}\right)\nonumber\\
&+\left( \begin{array}{cc}
                     -\mu-h\s+U n_{i\bar{\s}} & \s \D_i \\
                     \s {\D_i}^\ast & \mu-h\s-U n_{i\bar{\s}} 
                   \end{array}\right)
  \left( \begin{array}{c}
                     u_{mi\s}\\
                     v_{mi\bar{\s}}
                   \end{array}\right),
\end{align}
where $E_{m\s}$ is the eigenenergy specified by an integer $m=1,\cdots,N$ and spin $\s$.
Electron density $n_{i\s}\equiv\left< \hat{c}_{i\s}^\dagger \hat{c}_{i\s}^{\phantom {\dagger}}\right>$ and the site-dependent superconducting order parameter 
$\D_{i}\equiv U\left< \hat{c}_{i\ua}^{\phantom {\dagger}}\hat{c}_{i\da}^{\phantom {\dagger}}\right>$ are determined self-consistently as
\begin{align}
n_{i\s}=\sum_{m}\left[ |u_{mi\s}|^2 f(E_{m\s}) +|v_{mi\s}|^2\{1-f(E_{m\bar{\s}})\} \right]
\end{align}
and
\begin{align}
\D_{i}=U\sum_{m}\left[ u_{mi\da}v_{mi\ua}^\ast f(E_{m\s}) +u_{mi\ua}v_{mi\da}^\ast \{1-f(E_{m\bar{\s}})\} \right].
\end{align}
Due to the mean-field nature of the BdG theory, a superconductivity occurs in two dimensions, which should however be interpreted as a quasi-two-dimensional system where three dimensionality suppressing long-range fluctuations somehow comes in.

Figure \ref{fig:op-h}(a) plots the calculated $\D_i$ against the magnetic field $h$ for $\nb=0.5$ (quater filling). 
Since there are 1142 inequivalent sites in the 11006-site cluster, 1142 points are plotted at each $h$. 
We find that the behavior qualitatively changes from $T=0.3$ to $T=0.01$. 
At $T=0.3$, $\D_i$'s decrease as $h$ increases, with keeping the same sign for all the sites, and continuously vanish above $h_{c2} \simeq 0.33$.
At $T=0.01$, on the other hand, $\D_i$'s are nearly unchanged up to $h=0.5$, above which $\D_i$'s suddenly acquire site-dependent signs. 
Their amplitudes diminish as $h$ is increased further, at least up to $h\simeq 0.7$. 
Interestingly, $\D_i$ averaged over the sites is always nearly zero in this regime, showing a nearly equal contribution from positive and negative $\D_i$'s.
For $0.7<h<0.85$ several different spatial patterns seem to be competing, as $\D_i$'s do not monotonically decrease with $h$.
However, in this region, substantial values of $\D_i$ are found only around the edges of the cluster so that we do not explore this region in a further detail.
Eventually, above $h_{c2}\simeq 0.85$ the normal phase appears. 
Note that, unlike in periodic systems, the superconducting state in quasiperiodic systems is always inhomogeneous with respect to the {\it amplitude} of $\D_i$ even without magnetic field  \cite{sakai17} while its phase (sign) is uniform.
The finding here is an inhomogeneous {\it sign} structure in the superconducting state, which emerges under the magnetic field.

Corresponding to the abrupt change of $\D_i$ at $T=0.01$ and $h=0.52$, the magnetization $m_i\equiv n_{i\ua}-n_{i\da}$ also shows an abrupt change [Fig.~\ref{fig:op-h}(b)]: It is nearly zero for $h<0.5$ and jumps to finite values at $h=0.52$.
This abrupt change is not seen for $T=0.3$, where $m_i$'s gradually increase from $h=0$.

Figures \ref{fig:op-site}(a)-(d) are real-space maps of $\D_i$, where the coordinate of site $i$ is represented by $\Vec{r}_i=(x_i,y_i)$ with setting the length of each edge of the rhombuses to be 1.
The red (blue) points represent a positive (negative) value of $\D_i$ while gray points do nearly zero value. 
At $h=0$ [Fig.~\ref{fig:op-site}(a)], $\D_i$ is positive at every sites and this uniform-sign structure persists up to $h=0.5$, as is seen in Fig.~\ref{fig:op-h}(a).
Then, at $h=0.52$, the spatial pattern changes abruptly, showing a fractallike oscillation with keeping the five-fold rotational symmetry [Fig.~\ref{fig:op-site}(b)].
Figures \ref{fig:op-site}(c) and (d) show the pattern at different $\nb$'s, where $h$ is chosen to the value just above the transition point from the uniform-sign phase to the alternating-sign phase.
In the gray region (where $\D_i \simeq 0$), the magnetization $m_i$ is enhanced as is seen in Figs.~\ref{fig:op-site}(f)-(h), indicating that the magnetization energy compensates the loss of condensation energy at the nodes of $\D_i$, analogously to the FFLO state in periodic systems.
This sign structure of $\D_i$ changes with $\nb$ and $h$: For smaller (larger) $\nb$  and $h$, less (more) nodes appear [Fig.~\ref{fig:op-site}(c)(d)]. 
This will be reasonable because the magnetization energy increases with $\nb$ and $h$.
The sign structure changes with $T$, too (Supprementary Material).

\begin{figure}[tb]
\center{
\includegraphics[width=0.48\textwidth]{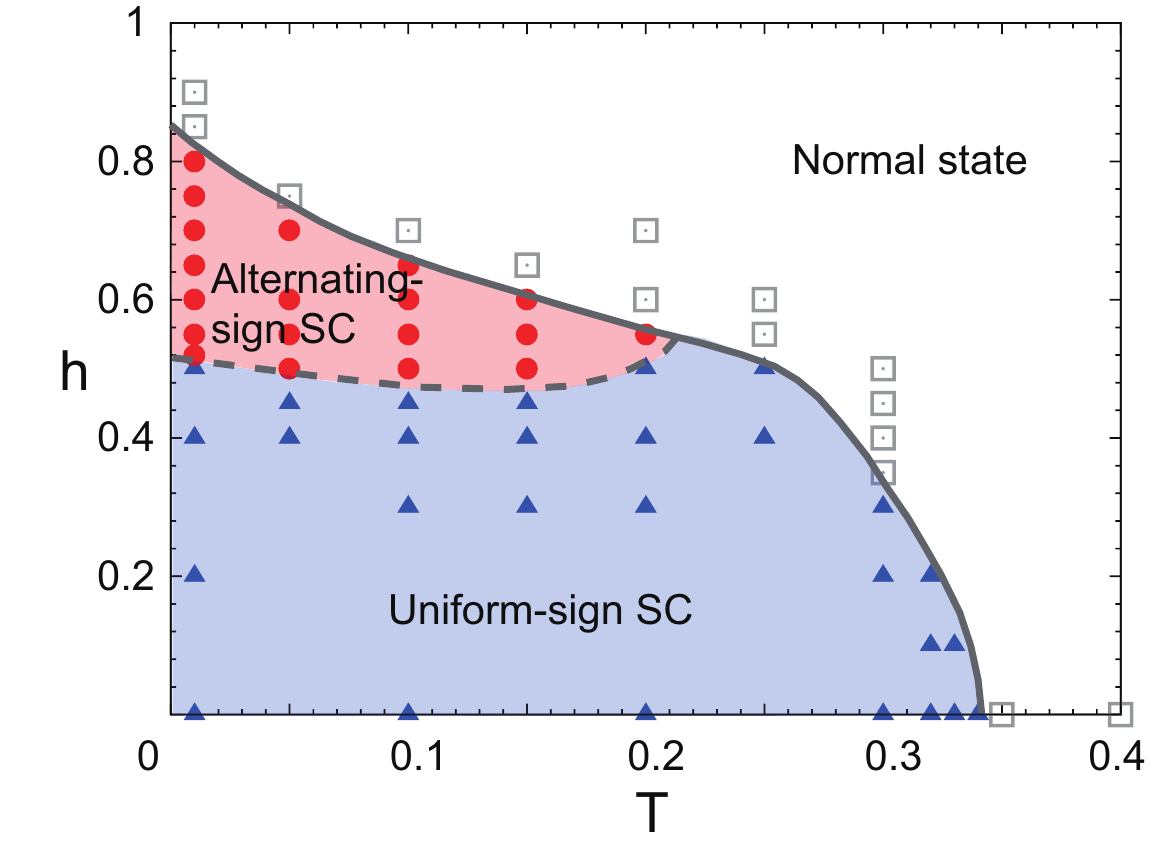}}
\caption{Phase diagram against temperature $T$ and magnetic field $h$, calculated for $\nb=0.5$, $U=-3$ and $N=11006$.
Squares, circles, and triangles denote the normal state, alternating-sign superconducting state, and uniform-sign superconducting state, respectively, where we regard the results to be superconducting when $\max_i|{\D_i}/U|>0.001$.}
\label{fig:pd}
\end{figure}

Figure \ref{fig:pd} shows the phase diagram against $h$ and $T$.
The alternating-sign superconducting state appears only in a relatively low-$T$ regime at high $h$ just below the critical field. 
The phase diagram bears a resemblance to those obtained for periodic systems \cite{takada69,burkhardt94,shimahara94,baarsma16,tylutki18} if we homologize the alternating-sign superconducting state with the FFLO phase in the periodic system.
In Supplementary Material, we investigate the dependence of the phases on other parameters, $\nb$ and $U$, too.


\begin{figure}[tb]
\center{
\includegraphics[width=0.48\textwidth]{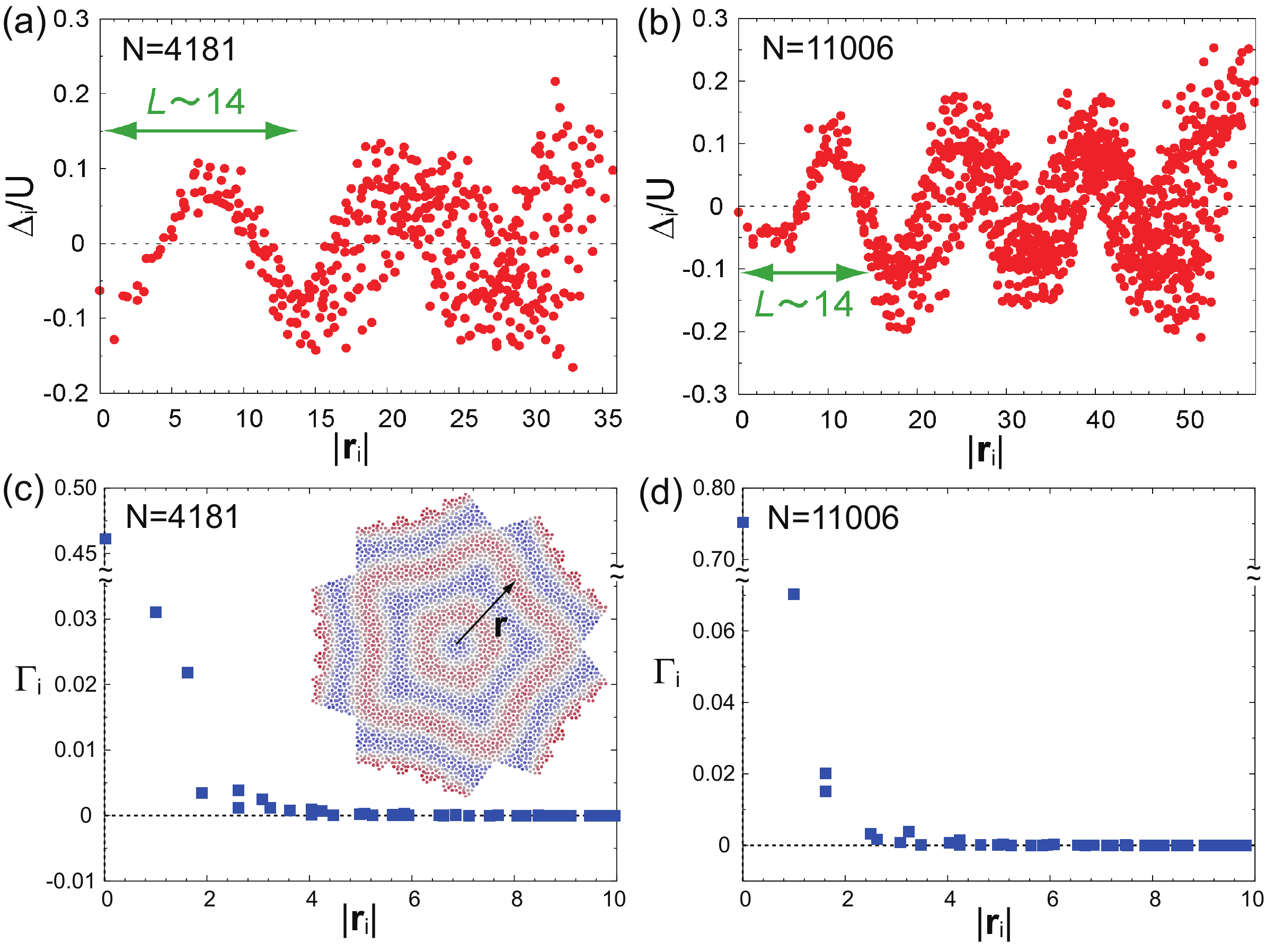}}
\caption{(a) (b) Onsite order parameter $\D_i$ plotted against the Euclidean distance $|\Vec{r}_i|=\sqrt{x_i^2+y_i^2}$ measured from the central site for $N=4181$ and $11006$, respectively. $\nb=0.5$, $T=0.01$ and $h=0.52$ are used. Green arrows denote the ``period" of the oscillation.
(c) (d) Corresponding $\G_i$ for the normal-state solution, plotted against $|\Vec{r}_i|$.
Inset to panel (c) shows the spatial pattern of $\D_i$ for $N=4181$.
}
\label{fig:op-r}
\end{figure}

The $h$-$T$ phase diagram is similar between $N=11006$ (Fig.~\ref{fig:pd}) and $N=4181$ (Fig.~\ref{fig:pd7} in Supplementary Material), showing that the result does not strongly depend on the size of the Penrose cluster. 
Moreover, in Figs.~\ref{fig:op-r}(a) and \ref{fig:op-r}(b), we compare the spatial structure of $\D_i$ between $N=4181$ and $11006$ for $\nb=0.5$, $T=0.01$ and $h=0.52$, where we plot $\D_i/U$ against the Euclidean distance $|\Vec{r}_i|=\sqrt{x_i^2+y_i^2}$ measured from the central site ($\Vec{r}_i=0$).
Although the geometry around the center differs between $N=4181$ and $11006$ clusters[, which is presumably a reason for a small difference (rotation) in the spatial pattern of $\D_i$ between Fig.~\ref{fig:op-site}(b) and the inset to Fig.~\ref{fig:op-r}(c) though we do not discuss this slight difference in a further detail], we find that the ``period" of the oscillation of $\D_i$ is almost the same between the two clusters: 
The green arrows in Figs.~\ref{fig:op-r}(a) and \ref{fig:op-r}(b) show the ``period" of about 14 in both cases.
This indicates that the ``period" is intrinsic to the Penrose structure rather than the boundary effect.

In order to understand the origin of this period, we first review a discussion about the FFLO state: 
In periodic systems at sufficiently low temperatures, the instability of the normal state to the Cooper pairing with total momentum $\Vec{q}$ is approximated by 
\begin{align}
T\int d^2\Vec{k} \sum_{n\s} G_{\s}(\Vec{k}+\Vec{q},i\w_n)G_{\bar{\s}}(-\Vec{k},i\w_n),\label{eq:gg}
\end{align}
where $G_{\s}(\Vec{k},i\w_n)$ is the spin-$\s$ component of the one-particle Green's function at momentum $\Vec{k}$ and the Matsubara frequency $\w_n\equiv(2n+1)\pi T$ in the normal state.
To make a connection to quasiperiodic systems, which lack a well-defined momentum, we substitute $G_{\s}(\Vec{k},i\w_n)$ with $\frac{1}{N}\sum_{j}G_{\s}(\Vec{r}_{j},i\w_n)e^{i\Vec{k}\cdot \Vec{r}_j}$, obtaining
\begin{align}
\frac{T}{N} \sum_{j n \s} G_{\s}(\Vec{r}_{j},i\w_n)G_{\bar{\s}}(\Vec{r}_j,i\w_n) e^{i\Vec{q}\cdot\Vec{r}_j}.\label{eq:gg2}
\end{align}
The momentum $\Vec{q}$ characteristic of the FFLO state is determined to maximize this quantity, with compromising between the condensation and magnetization energies.

In line with the above interpretation, we calculate a quantity 
$\G_{j}\equiv T \sum_{n\s} G_{\s}(\Vec{r}_{j},i\w_n)G_{\bar{\s}}(\Vec{r}_j,i\w_n)$ for the Penrose structure and plot $\G_{j}$ against $|\Vec{r}_j|$ in Figs.~\ref{fig:op-r}(c) and \ref{fig:op-r}(d).
Here, we have used the parameters for the alternating-sign pairing state of Fig.~\ref{fig:op-site}(b) while restricting a solution to the normal state to see its instability. 
We see that $\G_j$ is substantial only for $|\Vec{r}_j| < r_c\sim 3$ in both cases of $N$. 
This means that the instability is relatively large when the ``period" $L$ [which equals $\frac{2\pi}{|\Vec{q}|}$ in Eq.~(\ref{eq:gg2})] is chosen as $L> 4r_c$ (so that $\cos \frac{2\pi r}{L}>0$ for $-r_c<r<r_c$).
On the other hand, the energy gain due to the magnetization increases as the nodes of $\D_i$ increase, i.e., $L$ decreases. 
Then, as a compromise between the condensation and magnetization energies, $L$ will be chosen to be around $4r_c \sim 12$, explaining the period ($\sim 14$) observed in Figs.~\ref{fig:op-r}(a) and \ref{fig:op-r}(b).

These results show that the alternating-sign superconducting state emerges by a mechanism similar to FFLO. However, the existence of the alternating-sign superconducting state is still nontrivial in quasiperiodic systems because the FFLO state, which involves non-time-reversal pairs, is known to be unstable against nonmagnetic impurities \cite{aslamazov69,agterberg01}:
While quasiperiodic potentials often act as a random potential for electrons, this is not the present case: The electron system finds a way to keep a coherence even under the quasiperiodic potential, by self-organizing the sign structure compatible with the Penrose structure.

On the other hand, the quasiperiodic structure may still work as a random potential for the cyclotron motion of electrons.
If this is the case, it will suppress the orbital pair-breaking effect even when the magnetic-field direction is deviated from the plane.
Although it is known that the orbital pair-breaking effects are harmful to the FFLO state in periodic systems \cite{gruenberg66,buzdin96},
if the above argument is true, the quasicrystalline superconductors under magnetic field may offer a precious platform of this exotic type of superconductivity.

However, it is not clear at present whether this exotic superconductivity is realized in the recently-discovered first example of superconducting quasicrystal \cite{kamiya18}.
This is partly because its $\Tc$ is rather low $(\simeq 50\rm{mK})$, indicating a small effective attraction between the electrons: 
As Fig.~\ref{fig:pdu} in Supplementary Material shows, the area of the alternating-sign superconducting state shrinks with decreasing $|U|$.
In addition, as we can see in Fig.~\ref{fig:pd}, the alternating-sign superconductivity emerges only at relatively low temperatures compared to $\Tc$ of $h=0$, which further makes a detection difficult.
In any case, as the first theoretical study of the quasicrystalline superconductors under magnetic field, the present results pose an interesting new possibility of exotic pairs in these systems.


In summary, we have studied the effect of magnetic field on quasiperiodic superconductors.
We find an exotic superconducting state with spatially sign-alternating order parameter at high magnetic fields just below the critical field. 
The state is analogous to the FFLO state in periodic systems while its existence and the spatial pattern conforming to the quasiperiodic structure are nontrivial. 
We have revealed that the exotic superconductivity occurs as a consequence of the competition between magnetization and condensation energies and given an explanation on the origin of the spatial pattern.

\begin{acknowledgments}
We thank S. Hoshino for a valuable discussion.
This work was supported by JSPS KAKENHI Grant No. JP26800179 and JP16H06345.
\end{acknowledgments}

\bibliography{ref}

\pagebreak

\section{Supplementary Material}
\subsection{Results for 4181 sites}
\begin{figure}[H]
\center{
\includegraphics[width=0.45\textwidth]{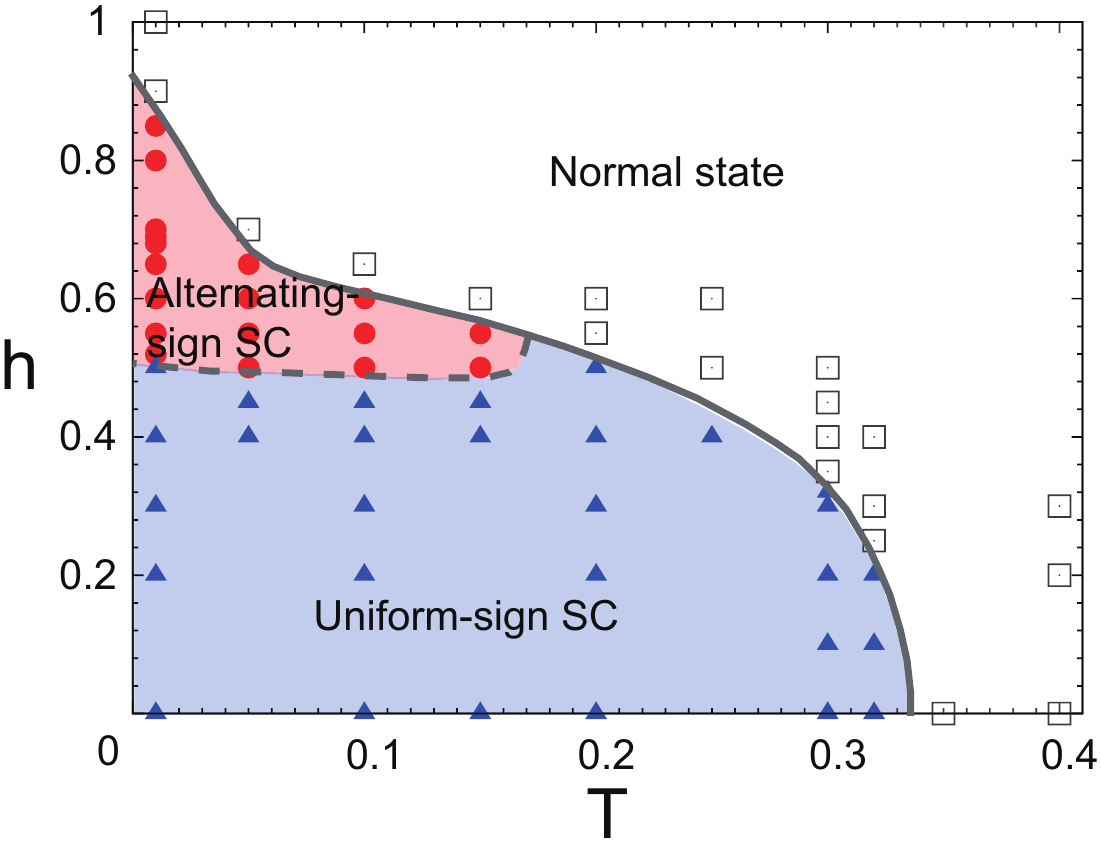}}
\caption{Phase diagram against $T$ and $h$ for $\nb=0.5$ and $U=-3$, calculated for the 4181-site cluster.
Squares, circles, and triangles denote the normal state, alternating-sign superconducting state, and uniform-sign superconducting state, respectively, where we regard the results to be superconducting when $\max_i|{\D_i}/U|>0.001$.}
\label{fig:pd7}
\end{figure}

Here we present results with a smaller size ($N=4181$) of the Penrose-tiling cluster, exploring dependence on various model parameters.
Figure \ref{fig:pd7} shows the phase diagram against $T$ and $h$, calculated for $\nb=0.5$.
We see that the results are similar to Fig.~\ref{fig:pd} for $N=11006$, confirming that the phase diagram is well converged against the cluster size.

\begin{figure}[H]
\center{
\includegraphics[width=0.45\textwidth]{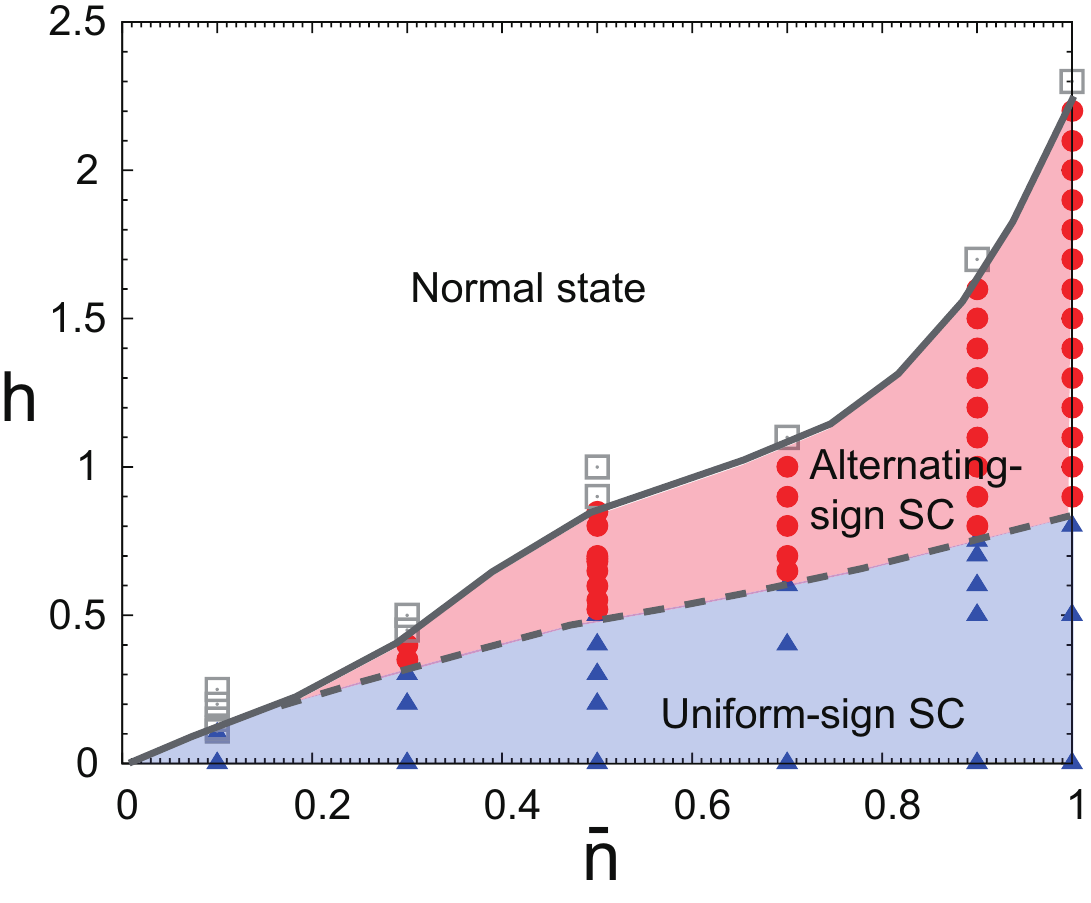}}
\caption{Phase diagram against $\nb$ and $h$ at $T=0.01$ and $U=-3$, calculated for the 4181-site cluster. The notations are the same as those used in Fig.~\ref{fig:pd7}.}
\label{fig:pdn}
\end{figure}

Figure \ref{fig:pdn} shows the phase diagram against $\nb$ and $h$ at $T=0.01$.
Note that the phase diagram is symmetric with respect to $\nb=1$ in the present model so that we have plotted it only for $\nb\leq 1$.
We see that, as $\nb$ increases, the superconducting region expands and accordingly the alternating-sign superconducting state occupies a wider region of $h$. For $\nb<0.2$ the alternating-sign superconducting state is not found at this temperature.

Figure \ref{fig:pdu} shows the phase diagram against $|U|$ and $h$ at $T=0.01$ and $\nb=0.5$.
We find that the area of the alternating-sign superconducting state shrinks as $|U|$ decreases.
For $|U|<2$ we do not find the alternating-sign superconductivity at this temperature.
Note that for a large value of $|U|$ (comparable to the bare ``bandwidth" $\sim 8.5t$), the present BdG approach may not be appropriate.

\subsection{Results at higher temperatures}
\begin{figure}[b]
\center{
\includegraphics[width=0.48\textwidth]{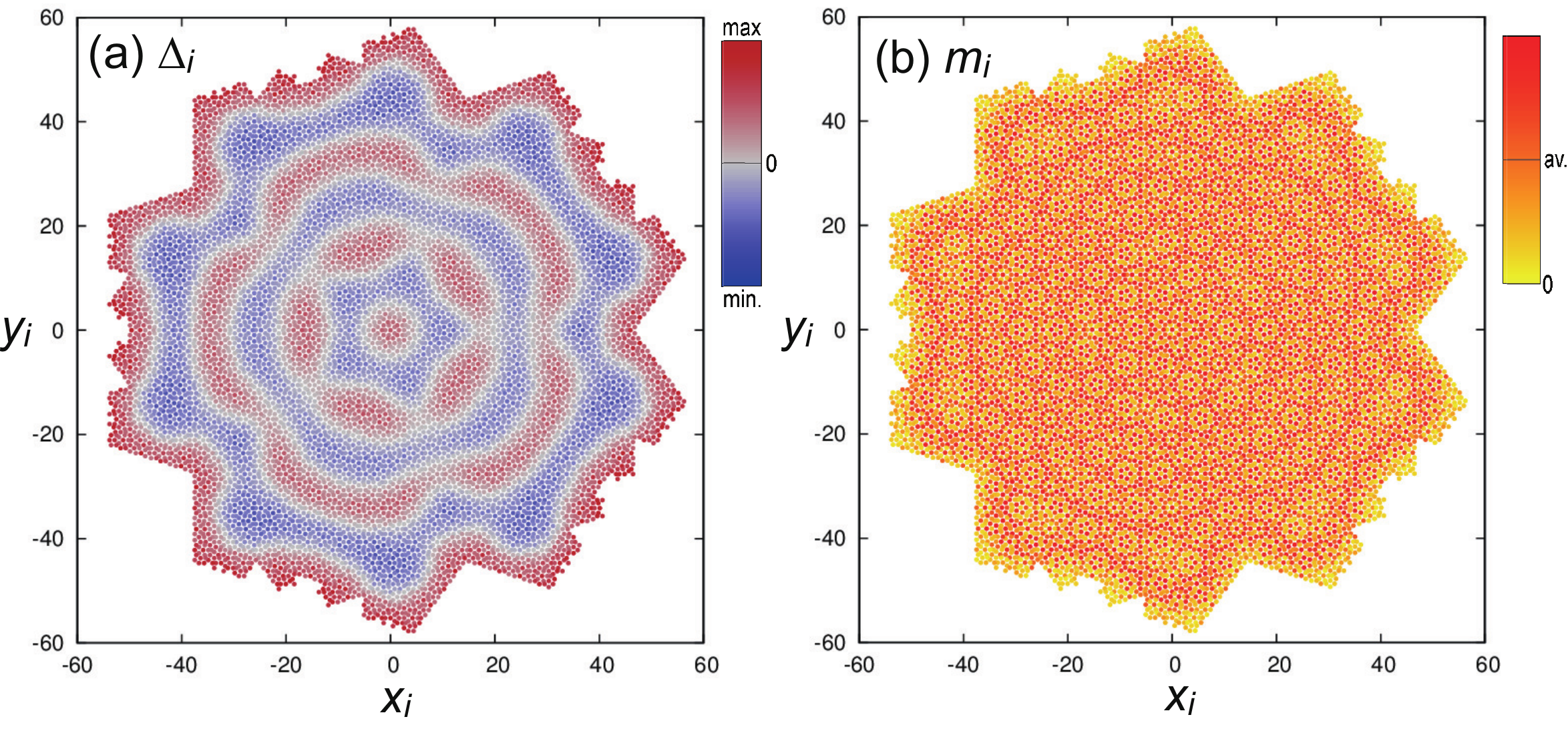}}
\caption{Spatial pattern of (a) $\D_i$ and (b) $m_i$ for $T=0.1$, $\nb=0.5$, $U=-3$, $h=0.5$ and $N=11006$.}
\label{fig:op-hight}
\end{figure}

Figure \ref{fig:op-hight} shows a real-space map of $\D_i$ and $m_i$ at a relatively high temperature $T=0.1$ for $\nb=0.5$ and $h=0.5$ in the alternating-sign superconducting phase.
Compared to Fig.~\ref{fig:op-site}(a) at a lower temperature $T=0.01$, Fig.~\ref{fig:op-hight}(a) shows a longer period of the sign oscillation presumably because of a thermal broadening of the fine structure.

\begin{figure}[H]
\center{
\includegraphics[width=0.45\textwidth]{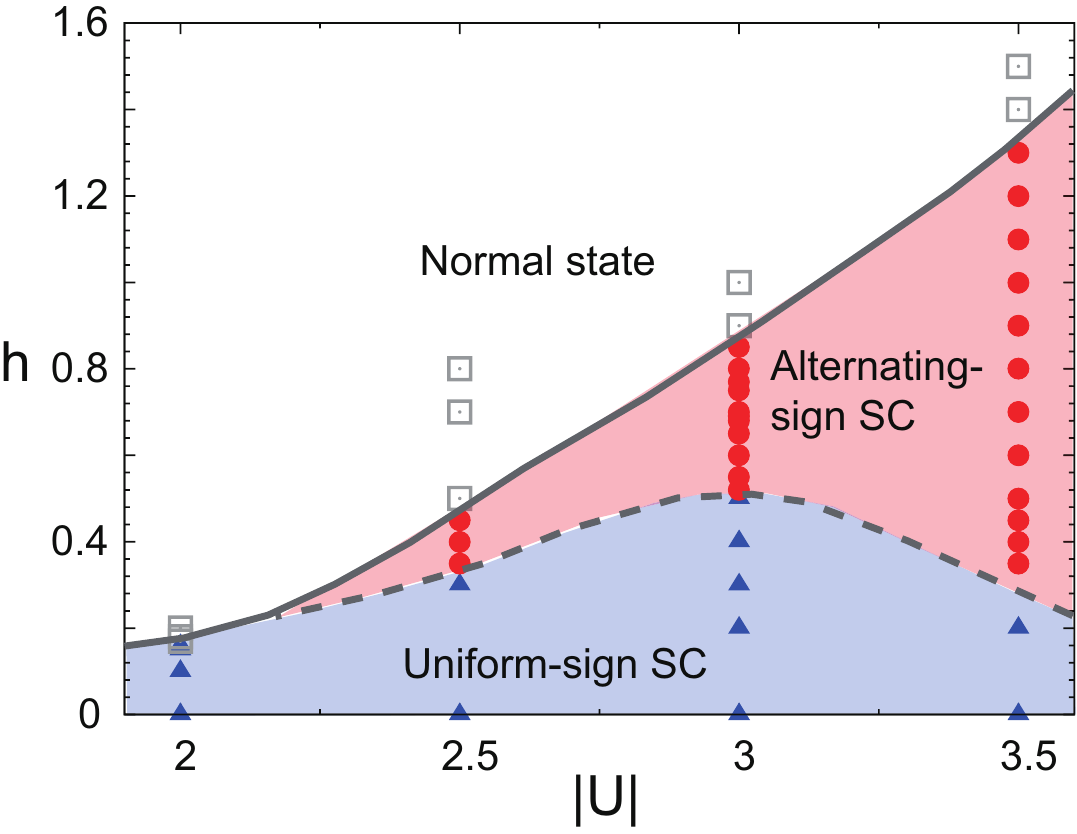}}
\caption{Phase diagram against $|U|$ and $h$ at $T=0.01$ and $\nb=0.5$, calculated for the 4181-site cluster.
The notations are the same as those used in Fig.~\ref{fig:pd7}. }
\label{fig:pdu}
\end{figure}

\end{document}